\documentclass{article}
\usepackage[T1]{fontenc}
\usepackage[latin9]{inputenc}
\usepackage[letterpaper]{geometry}
\usepackage{textcomp}
\usepackage{amstext}
\usepackage{graphicx}
\usepackage{esint}
\usepackage[authoryear]{natbib}

\makeatletter
\newcommand{\lyxaddress}[1]{
	\par {\raggedright #1
	\vspace{1.4em}
	\noindent\par}
}

\makeatother

\begin{document}
\title{Constraints on the latitudinal profile of Jupiter's deep jets}
\author{E. Galanti$^{1}$, Y. Kaspi$^{1}$, K. Duerl$^{1}$, L. Fletcher$^{2}$, A. P. Ingersolll$^{3}$,\\
         C. Li$^{4}$,   G. S. Orton$^{5}$, T. Guillot$^{6}$,  S. M. Levin$^{5}$, and S. J. Bolton$^{7}$
\\
(Geophysical Research Letters, submitted)}
\maketitle

\lyxaddress{\begin{center}
\textit{$^{1}$Department of Earth and Planetary Sciences, Weizmann
Institute of Science, Rehovot, Israel. }\\
\textit{$^{2}$School of Physics and Astronomy, University of Leicester, Leicester, UK}\\
\textit{$^{3}$California Institute of Technology, Pasadena, CA, USA}\\
\textit{$^{4}$Department of Climate and Space Sciences and Engineering, University of Michigan, Ann Arbor, MI, USA}\\
\textit{$^{5}$Jet Propulsion Laboratory, California Institute of Technology, Pasadena, USA}\\
\textit{$^{6}$Observatoire de la Cote d\textquoteright Azur, Nice, France}\\
\textit{$^{7}$Southwest Research Institute, San Antonio, Texas, TX, USA}
\par\end{center}}

\begin{abstract}
\textbf{The observed zonal winds at Jupiter's cloud tops have been shown to
be closely linked to the asymmetric part of the planet's measured
gravity field. However, other measurements suggest that in some latitudinal
regions the flow below the clouds might be somewhat different from
the observed cloud-level winds. Here we show, using both the symmetric
and asymmetric parts of the measured gravity field, that the observed
cloud-level wind profile between 25$^{\circ}$S and 25$^{\circ}$N
must extend unaltered to depths of thousands of kilometers. Poleward,
the midlatitude deep jets also contribute to the gravity signal, but
might differ somewhat from the cloud-level winds. We analyze the likelihood
of this difference and give bounds to its strength. We also find that
to match the gravity measurements, the winds must project inward in
the direction parallel to Jupiter's spin axis, and that their decay
inward should be in the radial direction.}
\end{abstract}

\section{Introduction{\normalsize{}\label{sec:Introduction}}}

The zonal (east-west) wind at Jupiter's cloud level dominate the atmospheric
circulation, and strongly relate to the observed cloud bands \citet{Fletcher2020a}.
The structure of the flow beneath the cloud level has been investigated
by several of the instruments on board the Juno spacecraft by means
of gravity, infrared and microwave measurements \cite{Bolton2017}.
Particularly, the gravity measurements were used to infer that the
winds extend down to roughly 3000~km, and that the main north-south
asymmetry in the cloud-level wind extends to these great depths \cite{Kaspi2018},
resulting in the substantial values of the odd gravity harmonics $J_{3}$,
$J_{5}$, $J_{7}$, and $J_{9}$. The excellent match between the
sign and value of the predicted odd harmonics using the cloud-level
wind \cite{Kaspi2013a} and the Juno gravity measurements \cite{Iess2018},
led to the inference that the wind profile at depth is similar to
that at the cloud level \cite{Kaspi2018,Kaspi2020}. Here, we revisit
in more detail the relation between the exact meridional profile of
the zonal flow and the gravity measurements, and study how much of
the cloud-level wind must be retained in order to match the gravity
measurements.

Since the gravity measurements are sensitive to mass distribution,
they are not very sensitive to the shallow levels (0.5-240~bar) probed
by Juno's microwave radiometer (MWR, \citealp{Janssen2017}), as the
density in this region is low compared to the deeper levels. Yet,
the gravity measurements have substantial implications on the MWR
region, since if the flow profile at depth (below the MWR region)
resembles that at the cloud level it is likely that the flow profile
within the MWR region is not very different. In such a case, where
the flow is barotropic, this implies via thermal wind balance that
latitudinal temperature gradients in the MWR region are small, which
has important implication to the MWR analysis of water and ammonia
distribution \cite{Li2017,Ingersoll2017,Li2020}. Thus, it is important
to determine how strong the gravity constraint on the temperature
distribution is, and what is its latitudinal dependence.

The determination of the zonal flow field at depth is based on the
measurements of the odd gravity harmonics, $J_{3}$, $J_{5}$, $J_{7}$,
and $J_{9}$, which are uniquely related to the flow field \cite{Kaspi2013a}.
Using only four numbers to determine a 2D (latitude and depth) field
poses a uniqueness challenge, and solutions that are unrelated to
the observed cloud-level wind  can be found \cite{Kong2018}, although
the origin of such internal flow structure, completely unrelated to
the cloud-level winds, is not clear. In addition, these solutions
require a flow of about 1~m~s$^{-1}$ at depth of 0.8 the radius
of Jupiter ($\sim$15,000~km), where the significant conductivity
\cite{Liu2008,Wicht2019a} is expected to dampen such strong flows
\cite{Cao2017a,Duer2019,Moore2019}. Recently, \citep{Galanti2021}
showed that the interaction of the flow with the magnetic field in
the semiconducting region can be used as an additional constraint
on the structure of the flow below the cloud level. With some modification
of the observed cloud-level wind, well within its uncertainty range
\cite{Tollefson2017}, a solution can be found that explains the
odd gravity harmonics and abides the magnetic field constraints.

All of the above mentioned studies assumed that if the internal flow
is related to the observed surface winds, it will manifest its entire
latitudinal profile. However, some evidence suggests that at some
latitudinal regions the flow below the clouds might be different from
the winds at the cloud level. The Galileo probe, entering the Jovian
atmosphere around planetocentric latitude 6.5$^{\circ}$N \cite{Orton1998},
measured winds that strengthened from 80~ms$^{-1}$ at the cloud
level to $\sim$160~ms$^{-1}$ at a depth of 4~bars, from where
it remains approximately constant until a depth of 20~bars where
the probe stopped transmitting data \cite{Atkinson1998}. Such a
baroclinic shear got further support in studies of equatorial hot
spots \cite{Li2006a,Choi2013}. Recently, \citet{Duer2020} showed
that the MWR measurements of brightness temperature correlate to the
zonal wind's latitudinal profile. They found that profiles differing
to a limited extent from the cloud-level can still be consistent with
both MWR and gravity. Emanating from the correlations between MWR
and the zonal winds, \citet{Fletcher2021} suggested that the winds
at some latitudes might strengthen from the cloud level to a depth
of 4-8~bars, i.e. not far from where water is expected to be condensing,
and only then begin to decay downward. Alternatively, based on stability
considerations, it was suggested that while westward jets are not
altered much with depth, the eastward jets might increase by 50-100\%
\cite{Dowling1995,Dowling2020}. 

Furthermore, in the \citet{Kaspi2018} and \citet{Galanti2021} studies,
the observed cloud-level wind has been assumed to be projected into
the planet interior along the direction parallel to the spin axis
of Jupiter, based on theoretical arguments \citep{Busse1970,Busse1976}
and 3D simulations of the flow in a Jovian-like planet \citep[e.g.,][]{Busse1994,Kaspi2009,Christensen2001,Heimpel2016}.
Theoretically this requires the flow to be nearly barotropic, which
is not necessarily the case, particularly when considering the 3D
nature of the planetary interior. Another assumption made is that
the flow decays in the radial direction. This was based on the reasoning
that any mechanism acting to decay the flow, such as the increasing
conductivity \citep{Cao2017a}, compressibility \citep{Kaspi2009},
or the existence of a stable layer \citep{Debras2019,Christensen2020},
will depend on pressure and temperature, which to first order are
a function of depth. However, if the internal flow is organized in
cylinders it might be the case that the mechanism acting to decay
it strengthens also in the direction parallel to the spin axis.

Here we investigate what can be learned about the issues discussed
above, based on the measured gravity field, considering both the symmetric
and asymmetric components of the gravity field measurements. We study
the ability to fit the gravity measurements with a cloud-level wind
that is limited to a specific latitudinal range, thus identifying
the regions where the observed cloud-level wind is likely to extend
deep, and the regions where the interior flow might differ (section~\ref{sec:The-latitudinal-sensitivity}).
We also examine whether a stronger wind at the 4-8 bar level is compatible
with the gravity measurements, and if the assumptions regarding the
relation of the internal flow to the cloud level can be relaxed (section~\ref{sec:Variants}).
Finally, we examine the latitudinal dependence of the wind-induced
gravity harmonics when magnetohydrodynamics considerations are used
as additional constraints (section~\ref{sec:magnetohydrodynamic}).

\section{Defining the cloud-level wind and possible internal flow structures\label{sec:Defining-the-cloud-level}}

\begin{figure}
\centering{}\includegraphics[scale=0.45]{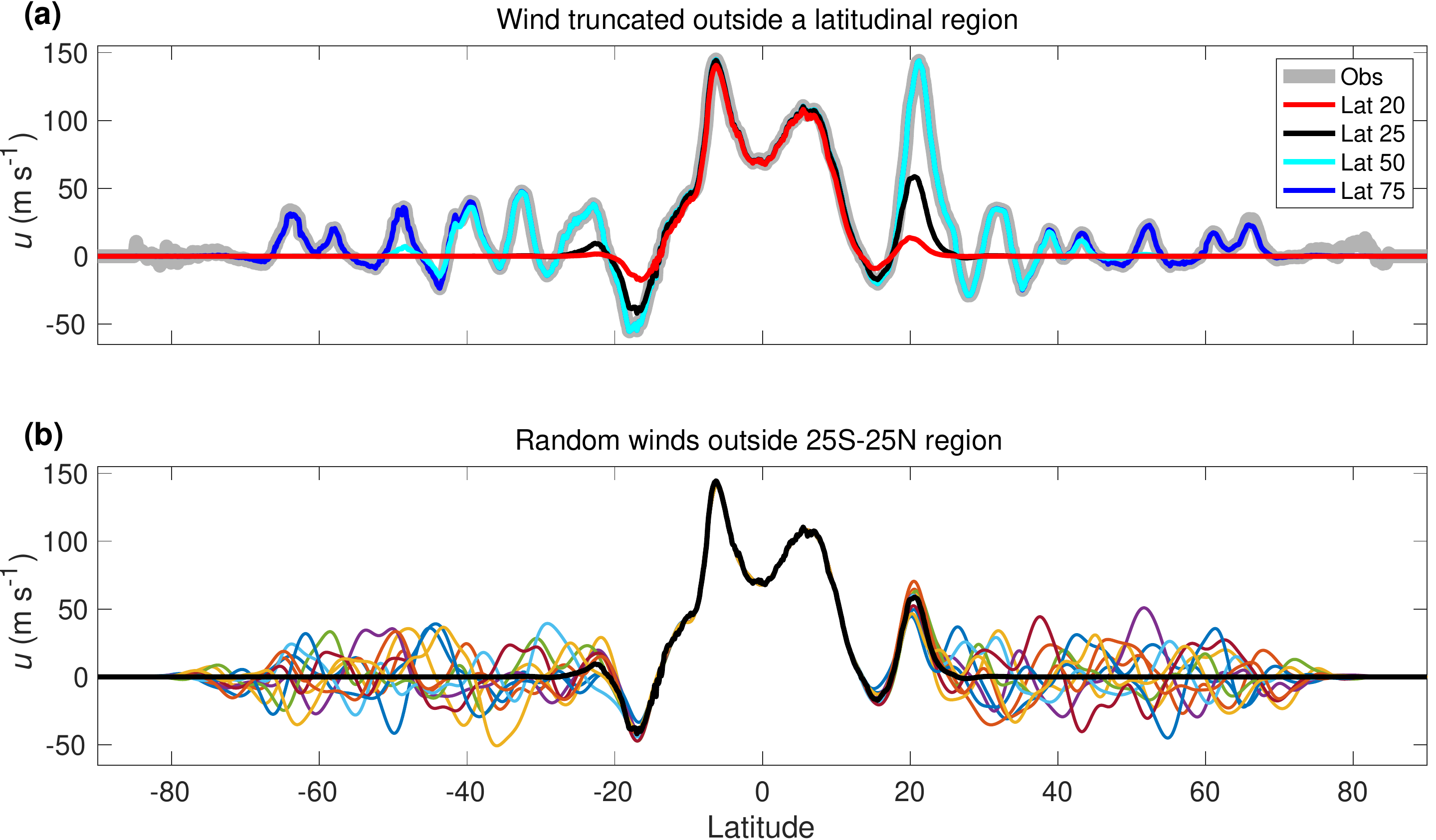}\caption{(a) The observed wind \citep{Tollefson2017} (gray), and variant examples
with the wind truncated poleward of the latitudes 20$^{\circ}$, 25$^{\circ}$,
50$^{\circ}$, and 75$^{\circ}$. (b) The case of wind truncated poleward
of the 25$^{\circ}$ latitude (black), along with examples of random
jets added in the truncated regions.\label{fig:Cloud-level-winds}}
\end{figure}

We examine several aspects of the flow structure that might influence
the ability to explain the gravity measurements. First, stemming from
the notion that at some latitudinal regions the flow below the cloud
level might differ from the observed, we set cases in which the cloud-level
wind is truncated at a specific latitude (Fig.~\ref{fig:Cloud-level-winds}a).
The truncation is done by applying a shifted hemispherically symmetric
hyperbolic tangent function with a transition width of 5$^{\circ}$,
to allow a smooth truncation of the wind from the observed flow. The
result is a wind profile that equatorward of the truncation latitude
is kept as in the cloud-top observations, and poleward decays quickly
to zero. We examine 18 cases with truncation latitudes $5^{\circ},10^{\circ},15^{\circ},...,90^{\circ}$.
Note that all of the cloud-level wind setups used in this study are
based on the analysis of the HST Jupiter images during Juno's PJ3
\citep{Tollefson2017}[, Figure~\ref{fig:Cloud-level-winds}a,
gray line], and that in all figures and calculations we use the planetocentric
latitude.

Next, we examine cases in which a different wind structure exists
poleward of the truncation latitude. As such, unknown wind structures
could possibly replace the observed cloud-level wind at shallow depths
of around 5-10~bars (e.g., as can be inferred from MWR, depending
on how microwave brightness temperatures are interpreted, see \citealt{Fletcher2021}).
For the purpose of the gravity calculation we treat these wind profiles
as if they replace the wind at the cloud level (the variation of the
wind between 1 and 10 bars has negligible effect on the induced gravity
field). The observed wind is truncated poleward of $25^{\circ}{\rm S}-25^{\circ}{\rm N}$,
and replaced with 1000 random wind structures that mimic the latitudinal
scale and strength of the observed winds (Fig.~\ref{fig:Cloud-level-winds}b).

\begin{figure}
\centering{}\includegraphics[scale=0.35]{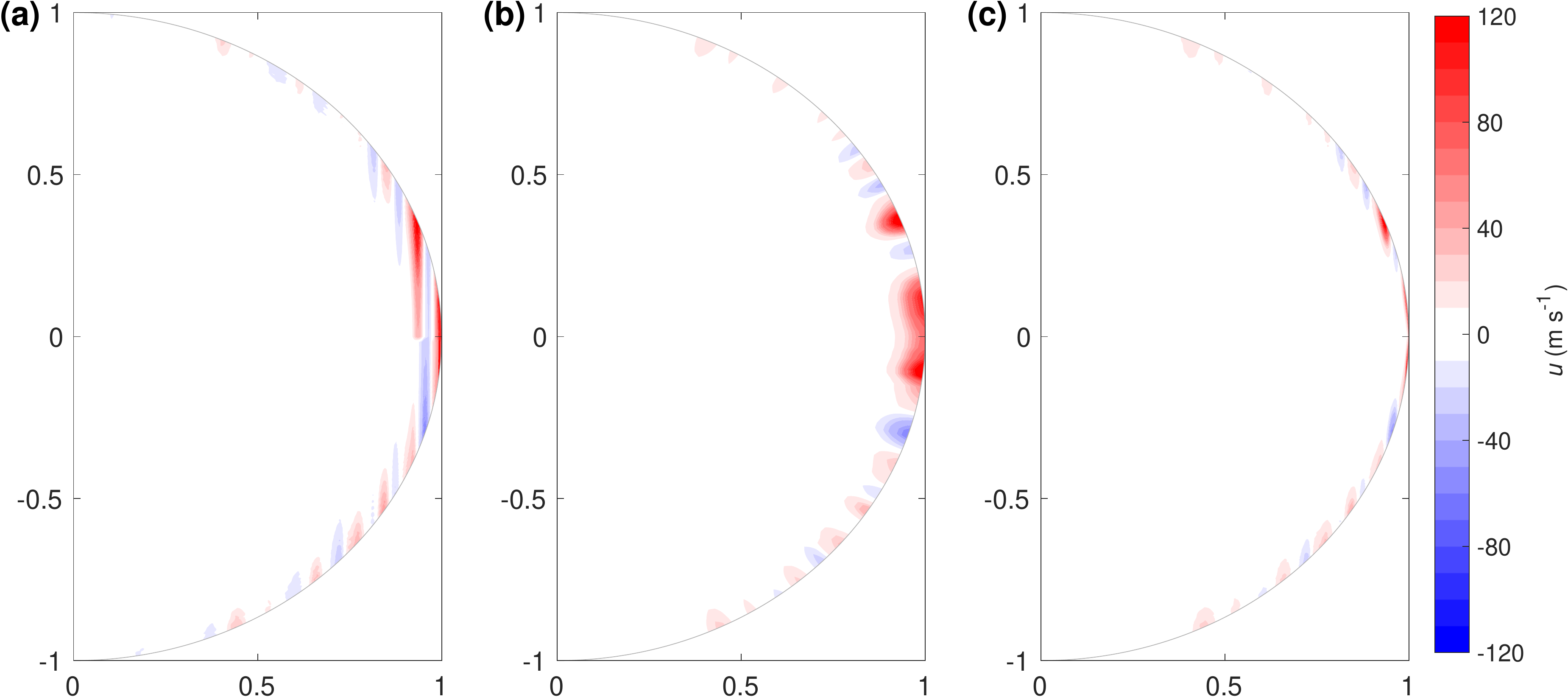}\caption{Options of cloud-level wind projection and decay profiles, shown for
an example of a sharp decay at a 3000~km distance from the surface.
(a) Projection in the direction parallel to the spin axis and decay
in the radial direction. (b) Projection and decay in the radial direction.
(c) Projection and decay in the direction parallel to the spin axis.\label{fig:Options-of-wind-decay}}
\end{figure}

The cloud-level wind profile is first projected inward in the direction
parallel to the spin axis \citep{Kaspi2010a}, and then made to decay
radially assuming a combination of functions (Fig.~\ref{fig:Options-of-wind-decay}a),
that allow a search for the optimal decay profile \cite[see also supporting information - SI]{Kaspi2018,Galanti2021}.
In addition, we examine two additional cases: a case in which the
cloud-level wind is both projected and decays in the radial direction
(Fig.~\ref{fig:Options-of-wind-decay}b), and a case in which the
wind is both projected and decays in the direction of the spin axis
(Fig.~\ref{fig:Options-of-wind-decay}c).

Given a zonal flow structure, thermal wind balance is used to calculate
an anomalous density structure associated with large-scale flow in
fast rotating gas giants. The density field is then integrated to
give the 1-bar gravity field in terms of the zonal gravity harmonics
\citep{Kaspi2010a}. Using an adjoint based optimization, a solution
for the flow structure is searched for, such that the model solution
for the gravity field is best fitted to the part of the measured gravity
field that can be attributed to the wind \citep{Galanti2016}. The
odd gravity harmonics are attributed solely to the wind, therefore
we use the Juno measured values $J_{3}=(-4.24\pm0.91)\times10^{-8}$,
$J_{5}=(-6.89\pm0.81)\times10^{-8}$, $J_{7}=(12.39\pm1.68)\times10^{-8}$,
and $J_{9}=(-10.58\pm4.35)\times10^{-8}$ \citep{Iess2018}. The lowest
even harmonics $J_{2}$ and $J_{4}$ are dominated by the planet's
density structure and shape and cannot be used in our analysis, but
interior models can give a reasonable estimate for the expected wind
contribution for the higher even harmonics $J_{6}$, $J_{8}$, and
$J_{10}$ \citep{Guillot2018}. Based on the Juno measurements and
the range of interior model solutions, the expected wind-induced even
harmonics are estimated as $\Delta J_{6}=1\times10^{-8}\pm(0.9+2)\times10^{-8}$,
$\Delta J_{8}=3.5\times10^{-8}\pm(2.46+0.5)\times10^{-8}$, and $\Delta J_{10}=-3\times10^{-8}\pm(6.94+0.25)\times10^{-8}$.
Note that the uncertainty associated with each even harmonic has contributions
from both the measurement and the range of interior model solutions
(first and second uncertainties, respectively). The large uncertainties
in the estimated wind-induced even harmonics suggest that our analysis
is limited to their order of magnitude and sign.

Finally, in order to isolate the latitudinal dependence from the general
ability to fit the gravity harmonics, we first optimize the cloud-level
wind so that the odd gravity harmonics are fitted perfectly \citep{Galanti2021}.
The modified wind is very similar to the observed (Fig.~S1), well
within the uncertainty of the cloud-level wind observation \citep{Tollefson2017},
therefore retaining all the observed latitudinal structure responsible
for the wind-induced gravity harmonics.

\section{The latitudinal sensitivity of the wind-induced gravity field\label{sec:The-latitudinal-sensitivity}}

\begin{figure}[t]
\centering{}\includegraphics[scale=0.7]{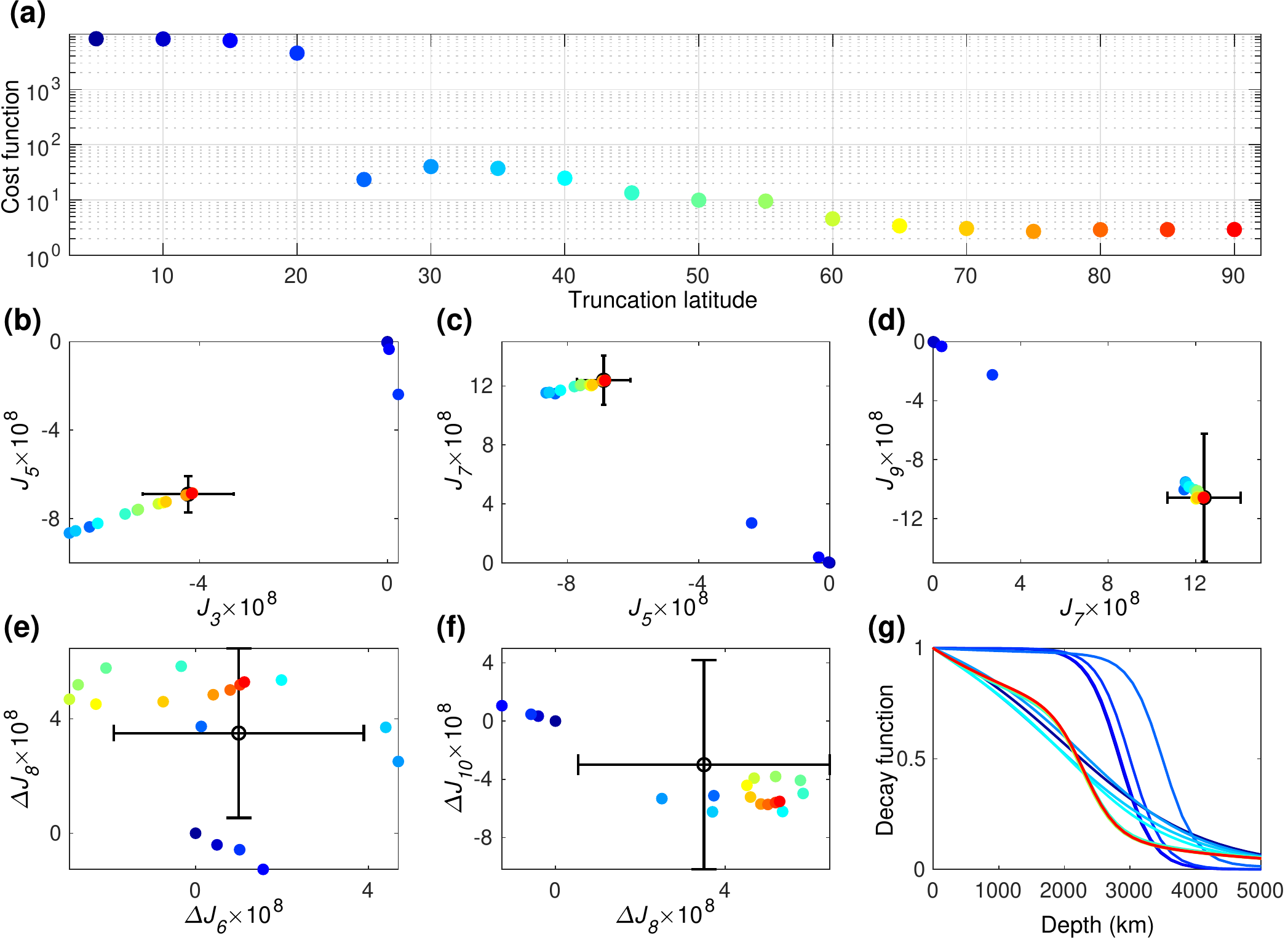}\caption{Latitude-dependent solutions as function of the truncation latitude.
(a) The overall fit of the model solution to the measurements (cost
function). Each case is assigned with a different color that is used
in the following panels, ranging from latitude 5$^{\circ}$ (blue)
to 90$^{\circ}$ (no truncation, red). (b-f) the solutions for the
different gravity harmonics (colors), and the measurement (black).
(g) the decay function associated with each solution.\label{fig:solutions-gravity-only}}
\end{figure}

We begin by analyzing the effect of the cloud-level wind latitudinal
truncation on the ability to explain the gravity harmonics. For each
wind setup, the internal flow structure is modified until the best
fit to the 4 odd harmonics and the 3 even harmonics is reached (Fig.~\ref{fig:solutions-gravity-only}).
The cost-function (Fig.~\ref{fig:solutions-gravity-only}a), a measure
for the overall difference between the measurements and the model
solution (see SI), reveals the contribution of each latitudinal region
to the solution. First, as expected, when the cloud-level wind is
retained at all latitudes, the solution for the odd harmonics is very
close to the measurements (Fig.~\ref{fig:solutions-gravity-only}b-d,
red dots). Importantly, the same optimal flow structure explains very
well the even harmonics (Fig.~\ref{fig:solutions-gravity-only}e-f,
red dots). This is additional evidence that the observed cloud-level
wind is dynamically related to the gravity field.

Examining the latitudinal dependence of the truncation, it is evident
that truncating the observed cloud-level wind closer to the equator
than $25^{\circ}{\rm S}-25^{\circ}{\rm N}$ prevents any flow structure
that could explain the gravity harmonics. It is most apparent in the
odd harmonics (Fig.~\ref{fig:solutions-gravity-only}b-d) where the
optimal solutions (dark blue circles) are close to zero and far from
the measured values. It is also the case for $\Delta J_{8}$ , but
for $\Delta J_{6}$ and $\Delta J_{10}$ the solutions are always
inside the uncertainty: in $\Delta J_{6}$ because the measured value
is very small, and in $\Delta J_{10}$ because the uncertainty is
very large. Considering the cloud-level wind profile (Fig.~\ref{fig:Cloud-level-winds}a,
black), it is not surprising that truncating the winds poleward of
$25^{\circ}{\rm S}-25^{\circ}{\rm N}$ makes the difference in the
solution, as this is where the positive (negative) jet in the northern
(southern) hemisphere are found, and project strongly on the low order
odd harmonics. Note that even a 5$^{\circ}$ difference (Fig.~\ref{fig:Cloud-level-winds}a,
red, truncation at $20^{\circ}{\rm S}-20^{\circ}{\rm N}$) prevents
a physical solution from being reached. Once these opposing jets are
included, the flow structure contains enough asymmetry to explain
very well $J_{7}$ and $J_{9}$ which have the largest values of the
odd harmonics.

However, with the $25^{\circ}{\rm S}-25^{\circ}{\rm N}$ truncation,
the model solutions for $J_{3}$ and $J_{5}$ are still outside the
measured uncertainty. Only when the influence of the zonal winds throughout
the $50^{\circ}{\rm S}-50^{\circ}{\rm N}$ range (Fig.~\ref{fig:Cloud-level-winds}a,
cyan) is included, then the lower odd harmonics can be explained with
the cloud-level wind profile. The optimal decay function for each
case (Fig.~\ref{fig:solutions-gravity-only}g), emphasize the robustness
of the solutions. When only the equatorial region is retained, the
optimization is trying (with no success) to include as much mass in
the region where the cloud-level wind is projected inward. But once
the winds at $25^{\circ}{\rm S}-25^{\circ}{\rm N}$ are included,
then the decay function of the wind settles on a similar profile,
with some small variations between the cases. Note that repeating
these experiments with the exact \citet{Tollefson2017} cloud-level
wind profile, does not change substantially the main results (Fig.~S2),
thus ensuring the robustness of the results.

The same methodology can be applied to a cloud-level wind that is
truncated equatorward of a latitudinal region (Fig.~S3). The analysis
shows that a wind truncated equatorward of a latitude larger than
$25^{\circ}{\rm S}-25^{\circ}{\rm N}$ does not allow a plausible
solution to be reached. Consistently with the above experiment, the
deep jets at $25^{\circ}{\rm S}-25^{\circ}{\rm N}$ are necessary
to fit gravity harmonics. Specifically, there is a gradual deterioration
of the solution in the truncation region of 0$^{\circ}$ to 20$^{\circ}$,
which is related solely to the even harmonics $\Delta J_{6}$ , $\Delta J_{8}$,
and $\Delta J_{10}$. Once the wind is truncated inside 10$^{\circ}$S-10$^{\circ}$N
the solution for $\Delta J_{6}$ and $\Delta J_{8}$ is outside the
uncertainty range, and $\Delta J_{10}$ moves further away from the
measurement. This is due to the strong eastward jets at 6$^{\circ}$S
and 6$^{\circ}$N.

\section{Variants of the flow structure\label{sec:Variants}}

\begin{figure}[t]
\centering{}\includegraphics[scale=0.5]{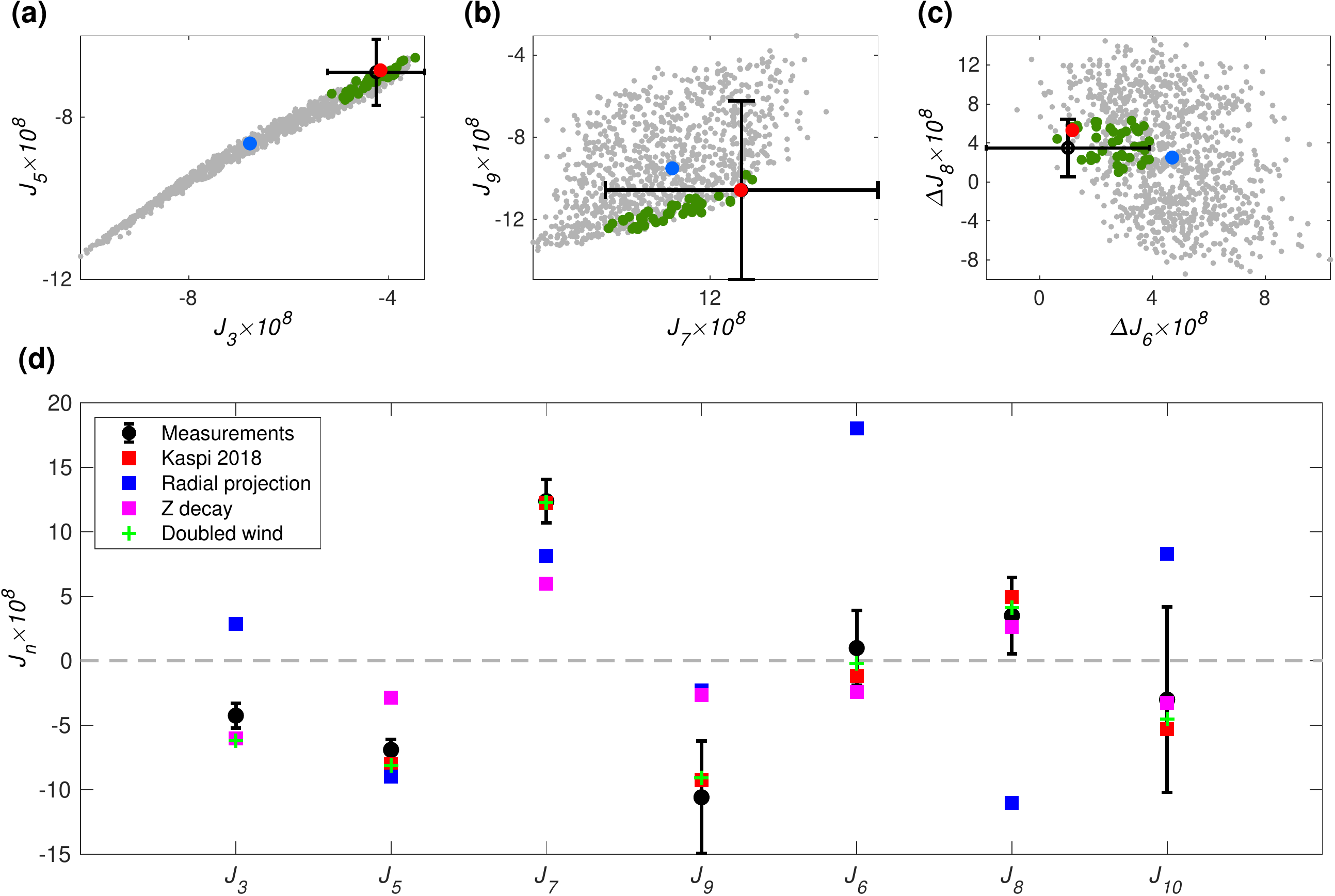}\caption{(a-c) Solutions with the cloud-level wind truncated poleward of $25^{\circ}{\rm S}-25^{\circ}{\rm N}$
and replaced with random jets there (Fig.~\ref{fig:Cloud-level-winds}b).
Shown are the solutions for 1000 random cases (gray), and within those
the solution which matches all the gravity harmonics (green). Also
shown are the solution with no random winds (blue, corresponding to
the 25$^{\circ}$ case in Fig.~\ref{fig:solutions-gravity-only}),
the solution with no truncation of the winds (red, corresponding to
the 90$^{\circ}$ case in Fig.~\ref{fig:solutions-gravity-only})
and the Juno measurements (black). (d) Solutions for cases with cloud-level
wind projected in the radial direction (blue; Fig.~\ref{fig:Options-of-wind-decay}b)
and wind decayed in the direction parallel to the spin axis (magenta;
Fig.~\ref{fig:Options-of-wind-decay}c), and a doubled cloud-level
wind (green). Also shown are the measurements (black), and the solution
with the unaltered cloud-level wind (red; \citealp{Kaspi2018}).\label{fig:solutions-random-others}}
\end{figure}

Next, we examine several variants to the wind setups. In section~\ref{sec:The-latitudinal-sensitivity}
we showed that the jets between 25$^{\circ}$S and 25$^{\circ}$N
are crucial for explaining the gravity harmonics, and therefore should
not differ much below the cloud level. However, in the regions where
the wind is truncated it should be examined whether a flow below the
cloud level that is completely different might still allow matching
the gravity harmonics. We therefore examine a case where the cloud-level
wind is truncated poleward of $25^{\circ}{\rm S}-25^{\circ}{\rm N}$,
and in the truncated regions random jets are added to simulate different
possible scenarios (Fig.~\ref{fig:Cloud-level-winds}b, see SI for
definition). The gravity harmonic solutions for 1000 different cases
is shown in Fig.~\ref{fig:solutions-random-others} (a-c). The largest
effect the random jets have is on $J_{3}$ and $J_{5}$, with considerable
effect also on the other odds and even harmonics. About 4\% of the
cases provide a good match to all the measurements (green), therefore
it is statistically possible that some combination of jets unseen
at the cloud level at the mid-latitudes, with amplitude of up to $\pm40$~m~s$^{-1}$,
are responsible for part of the gravity signal. Doubling (halving)
the random jets strength results in only 1.1\% (1.2\%) of the solutions
to fit the gravity measurement (SI, Fig.~S7), suggesting that if
alternative jets exists in the mid-latitudes, their amplitude should
be around $\pm40$~m~s$^{-1}$. These results are consistent with
\citet{Duer2020} who did a similar analysis, but taking the full
cloud level winds and showed that solutions differing from the cloud
level are possible but statistically unlikely ($\sim1\%$). 

Aside from modifications to the cloud-level wind, we also examine
cases in which the projection of the flow beneath the cloud level
is modified. For simplicity, we examine these cases with the observed
cloud-level wind spanning the full latitudinal range. Projecting the
wind radially and keeping the decay radial (Fig.~\ref{fig:Options-of-wind-decay}b),
we find that there is no plausible solution for flow structure under
these assumptions that would give a good fit to the gravity measurements
(Fig.~\ref{fig:solutions-random-others}d, blue). The best-fit model
solution for all $J_{n}$ is far from the measurements, well outside
their uncertainty range, and does not even match $J_{3}$ in sign.
Next, we consider a case in which the decay of the winds is in the
direction parallel to the spin axis (Fig.~\ref{fig:Options-of-wind-decay}c).
Here the optimal solution for the odd harmonics is far from the measured
values (Fig.~\ref{fig:solutions-random-others}d, magenta), while
for the even harmonics the solution is within the uncertainty range.
However, in this case the winds needs to be very deep, extending to
$\sim5000$~km, where the interaction with the magnetic field is
extremely strong \citep{Cao2017a,Galanti2017e,Galanti2021}. Finally,
following the suggestion that the cloud-level wind might get stronger
with depth before they decay \citep[e.g.,][]{Fletcher2021}, we conduct
an experiment in which we double the cloud-level wind. Interestingly,
a plausible solution can be achieved (Fig.~\ref{fig:solutions-random-others}d,
green crosses), with a decay profile similar to the \citet{Kaspi2018}
solution, but with the winds decaying more baroclinicaly in the upper
2000~km, and then decaying slower (Fig.~S6).

\begin{figure}[t]
\centering{}\includegraphics[scale=0.7]{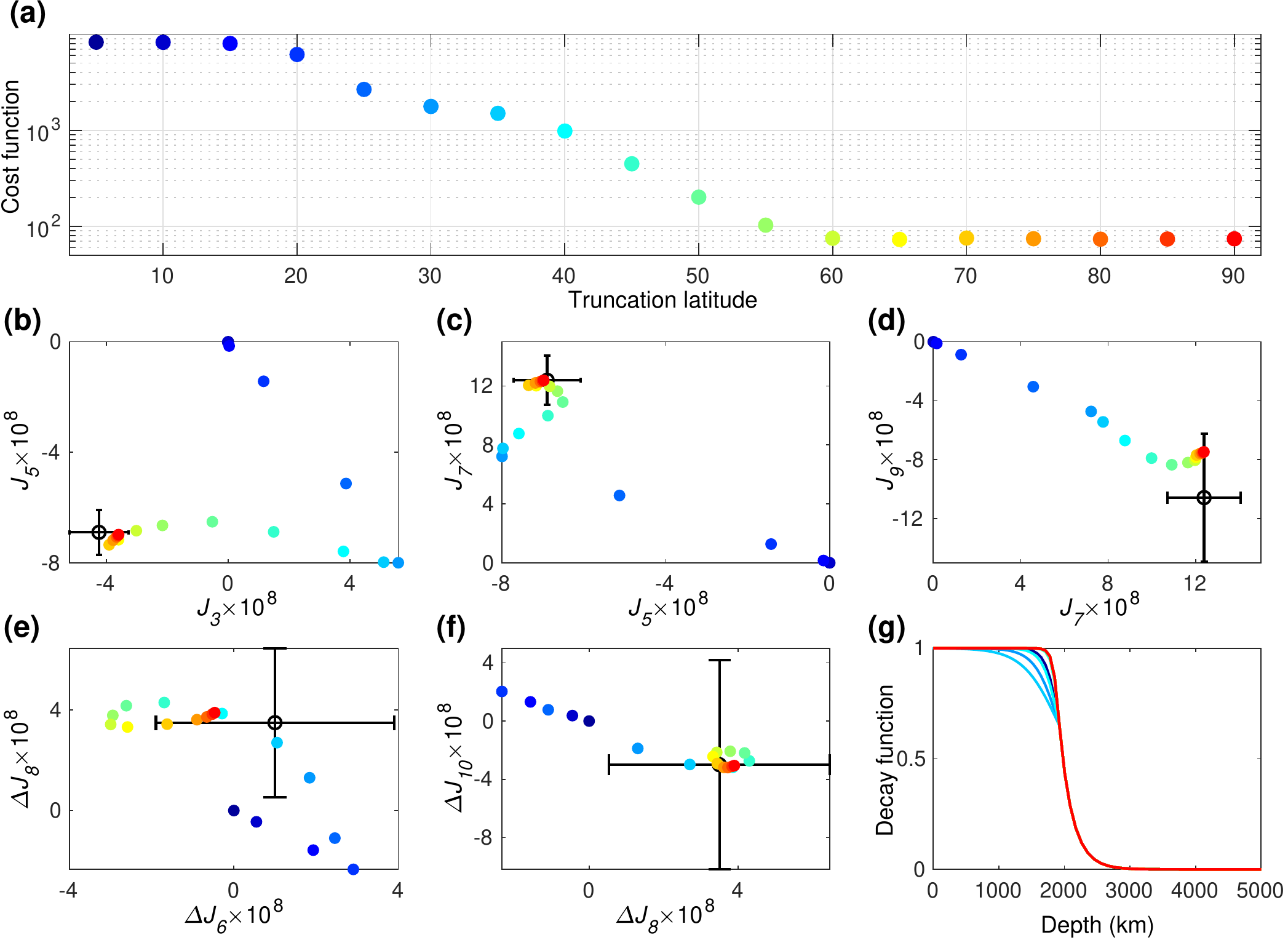}\caption{Same as Fig.~\ref{fig:solutions-gravity-only}, but for a case where
the flow profile in the semiconducting region is restricted to comply
with secular variations consideration.\label{fig:Solutions-magnetically-restricted}}
\end{figure}

\section{Adding magnetohydrodynamic constraints\label{sec:magnetohydrodynamic}}

In Jupiter, the increased conductivity with depth \citep[e.g.,][]{French2012,Wicht2019a}
suggests that the flow might be reduced to very small values in the
semiconducting region \citep[deeper than 2000 km,][]{Cao2017a}. Using
flow estimates in the semiconducting region based on past magnetic
secular variations \citep{Moore2019}, \citet{Galanti2021} gave a
revised wind decay profile that can explain both the gravity harmonics
and the constraints posed by the secular variations. We follow this
approach, setting the flow strength in the semiconducting region (deeper
than 2000~km, see \citealp{Galanti2021}) to be a sharp exponential
function (Fig.~\ref{fig:Solutions-magnetically-restricted}g, right
part). Given this inner profile of the decay function, the outer part
of the decay function can be searched for, together with the optimal
cloud-level wind, that will result in the best fit to the odd measured
gravity harmonics. The optimal cloud-level wind (Fig.~S1b) is very
similar to the observed wind, with deviations that are within the
uncertainties. 

Using the modified cloud-level wind, the shape of the decay function
in the outer neutral region is optimized to allow the best-fit to
the odd and even gravity harmonics (Fig.~\ref{fig:Solutions-magnetically-restricted}b-g).
In addition to the odd harmonics, which are expected to fit the measurements,
the model also fits very well the even harmonics, despite the limited
range of possible decay profiles in the outer region (Fig.~\ref{fig:Solutions-magnetically-restricted}g).
The latitudinal dependence reveals that the range of $50^{\circ}{\rm S}-50^{\circ}{\rm N}$
is needed in order to allow a good fit, especially for $J_{3}$ and
$J_{7}$. Similar to the case with gravity-only constraints, fitting
the even harmonics, as well as $J_{5}$ and $J_{9}$., requires mostly
the cloud-level wind inside the $25^{\circ}{\rm S}-25^{\circ}{\rm N}$
region. Thus, even when including the strong magnetic constraint,
the dominance of the $25^{\circ}{\rm S}-25^{\circ}{\rm N}$ region
remains robust.


\begin{thebibliography}{39}
\providecommand{\natexlab}[1]{#1}
\expandafter\ifx\csname urlstyle\endcsname\relax
  \providecommand{\doi}[1]{doi:\discretionary{}{}{}#1}\else
  \providecommand{\doi}{doi:\discretionary{}{}{}\begingroup
  \urlstyle{rm}\Url}\fi

\bibitem[{\textit{{Atkinson} et~al.}(1998)\textit{{Atkinson}, {Pollack}, and
  {Seiff}}}]{Atkinson1998}
{Atkinson}, D.~H., J.~B. {Pollack}, and A.~{Seiff}, The {G}alileo probe doppler
  wind experiment: Measurement of the deep zonal winds on {J}upiter, \textit{J.
  Geophys. Res.}, \textit{103}, 22,911--22,928, \doi{10.1029/98JE00060}, 1998.

\bibitem[{\textit{{Bolton} et~al.}(2017)}]{Bolton2017}
{Bolton}, S.~J., et~al., Jupiter's interior and deep atmosphere: The initial
  pole-to-pole passes with the {J}uno spacecraft, \textit{Science},
  \textit{356}, 821--825, \doi{10.1126/science.aal2108}, 2017.

\bibitem[{\textit{Busse}(1970)}]{Busse1970}
Busse, F.~H., Thermal instabilities in rapidly rotating systems, \textit{J.
  Fluid Mech.}, \textit{44}, 441--460, \doi{10.1017/S0022112070001921}, 1970.

\bibitem[{\textit{Busse}(1976)}]{Busse1976}
Busse, F.~H., A simple model of convection in the {J}ovian atmosphere,
  \textit{Icarus}, \textit{29}, 255--260, \doi{10.1016/0019-1035(76)90053-1},
  1976.

\bibitem[{\textit{Busse}(1994)}]{Busse1994}
Busse, F.~H., Convection driven zonal flows and vortices in the major planets,
  \textit{Chaos}, \textit{4}(2), 123--134, \doi{10.1063/1.165999}, 1994.

\bibitem[{\textit{{Cao} and {Stevenson}}(2017)}]{Cao2017a}
{Cao}, H., and D.~J. {Stevenson}, {Zonal flow magnetic field interaction in the
  semi-conducting region of giant planets}, \textit{Icarus}, \textit{296},
  59--72, \doi{10.1016/j.icarus.2017.05.015}, 2017.

\bibitem[{\textit{{Choi} et~al.}(2013)\textit{{Choi}, {Showman}, {Vasavada},
  and {Simon-Miller}}}]{Choi2013}
{Choi}, D.~S., A.~P. {Showman}, A.~R. {Vasavada}, and A.~A. {Simon-Miller},
  {Meteorology of Jupiter's equatorial hot spots and plumes from Cassini},
  \textit{Icarus}, \textit{223}(2), 832--843,
  \doi{10.1016/j.icarus.2013.02.001}, 2013.

\bibitem[{\textit{{Christensen}}(2001)}]{Christensen2001}
{Christensen}, U.~R., Zonal flow driven by deep convection in the major
  planets, \textit{Geophys. Res. Lett.}, \textit{28}, 2553--2556,
  \doi{10.1029/2000GL012643}, 2001.

\bibitem[{\textit{{Christensen} et~al.}(2020)\textit{{Christensen}, {Wicht},
  and {Dietrich}}}]{Christensen2020}
{Christensen}, U.~R., J.~{Wicht}, and W.~{Dietrich}, {Mechanisms for Limiting
  the Depth of Zonal Winds in the Gas Giant Planets}, \textit{Astrophys. J.},
  \textit{890}(1), 61, \doi{10.3847/1538-4357/ab698c}, 2020.

\bibitem[{\textit{{Debras} and {Chabrier}}(2019)}]{Debras2019}
{Debras}, F., and G.~{Chabrier}, New models of {J}upiter in the context of
  {J}uno and {G}alileo, \textit{Astrophys. J.}, \textit{872}, 100--,
  \doi{10.3847/1538-4357/aaff65}, 2019.

\bibitem[{\textit{Dowling}(1995)}]{Dowling1995}
Dowling, T.~E., Estimate of {J}upiter's deep zonal-wind profile from
  {S}hoemaker-{L}evy 9 data and {A}rnold's second stability criterion,
  \textit{Icarus}, \textit{117}, 439--442, \doi{10.1006/icar.1995.1169}, 1995.

\bibitem[{\textit{{Dowling}}(2020)}]{Dowling2020}
{Dowling}, T.~E., {Jupiter-style Jet Stability}, \textit{The Planatary Science
  Journal}, \textit{1}, 6, \doi{10.3847/PSJ/ab789d}, 2020.

\bibitem[{\textit{{Duer} et~al.}(2019)\textit{{Duer}, {Galanti}, and
  {Kaspi}}}]{Duer2019}
{Duer}, K., E.~{Galanti}, and Y.~{Kaspi}, Analysis of {J}upiter's deep jets
  combining {J}uno gravity and time-varying magnetic field measurements,
  \textit{Astrophys. J. Let.}, \textit{879}(2), L22,
  \doi{10.3847/2041-8213/ab288e}, 2019.

\bibitem[{\textit{{Duer} et~al.}(2020)\textit{{Duer}, {Galanti}, and
  {Kaspi}}}]{Duer2020}
{Duer}, K., E.~{Galanti}, and Y.~{Kaspi}, {The Range of Jupiter's Flow
  Structures that Fit the Juno Asymmetric Gravity Measurements}, \textit{J.
  Geophys. Res. (Planets)}, \textit{125}(8), e06292,
  \doi{10.1029/2019JE006292}, 2020.

\bibitem[{\textit{{Fletcher} et~al.}(2020)\textit{{Fletcher}, {Kaspi},
  {Guillot}, and {Showman}}}]{Fletcher2020a}
{Fletcher}, L.~N., Y.~{Kaspi}, T.~{Guillot}, and A.~P. {Showman}, {How Well Do
  We Understand the Belt/Zone Circulation of Giant Planet Atmospheres?},
  \textit{Space Sci. Rev.}, \textit{216}(2), 30,
  \doi{10.1007/s11214-019-0631-9}, 2020.

\bibitem[{\textit{{Fletcher} et~al.}(2021)}]{Fletcher2021}
{Fletcher}, L.~N., et~al., Jupiter's temperate belt/zone contrasts revealed at
  depth by {J}uno microwave observations, \textit{submitted}, 2021.

\bibitem[{\textit{{French} et~al.}(2012)\textit{{French}, {Becker}, {Lorenzen},
  {Nettelmann}, {Bethkenhagen}, {Wicht}, and {Redmer}}}]{French2012}
{French}, M., A.~{Becker}, W.~{Lorenzen}, N.~{Nettelmann}, M.~{Bethkenhagen},
  J.~{Wicht}, and R.~{Redmer}, Ab initio simulations for material properties
  along the {J}upiter adiabat, \textit{Astrophys. J. Sup.}, \textit{202}(1), 5,
  \doi{10.1088/0067-0049/202/1/5}, 2012.

\bibitem[{\textit{{Galanti} and {Kaspi}}(2016)}]{Galanti2016}
{Galanti}, E., and Y.~{Kaspi}, {An Adjoint-based Method for the Inversion of
  the Juno and Cassini Gravity Measurements into Wind Fields},
  \textit{Astrophys. J.}, \textit{820}(2), 91,
  \doi{10.3847/0004-637X/820/2/91}, 2016.

\bibitem[{\textit{{Galanti} and {Kaspi}}(2021)}]{Galanti2021}
{Galanti}, E., and Y.~{Kaspi}, {Combined magnetic and gravity measurements
  probe the deep zonal flows of the gas giants}, \textit{MNRAS},
  \textit{501}(2), 2352--2362, \doi{10.1093/mnras/staa3722}, 2021.

\bibitem[{\textit{{Galanti} et~al.}(2017)\textit{{Galanti}, {Cao}, and
  {Kaspi}}}]{Galanti2017e}
{Galanti}, E., H.~{Cao}, and Y.~{Kaspi}, Constraining {J}upiter's internal
  flows using {J}uno magnetic and gravity measurements, \textit{Geophys. Res.
  Lett.}, \textit{44}(16), 8173--8181, \doi{10.1002/2017GL074903}, 2017.

\bibitem[{\textit{{Guillot} et~al.}(2018)}]{Guillot2018}
{Guillot}, T., et~al., A suppression of differential rotation in {J}upiter's
  deep interior, \textit{Nature}, \textit{555}, 227--230,
  \doi{10.1038/nature25775}, 2018.

\bibitem[{\textit{{Heimpel} et~al.}(2016)\textit{{Heimpel}, {Gastine}, and
  {Wicht}}}]{Heimpel2016}
{Heimpel}, M., T.~{Gastine}, and J.~{Wicht}, Simulation of deep-seated zonal
  jets and shallow vortices in gas giant atmospheres, \textit{Nature
  Geoscience}, \textit{9}, 19--23, \doi{10.1038/ngeo2601}, 2016.

\bibitem[{\textit{{Iess} et~al.}(2018)}]{Iess2018}
{Iess}, L., et~al., Measurement of {J}upiter's asymmetric gravity field,
  \textit{Nature}, \textit{555}, 220--222, \doi{10.1038/nature25776}, 2018.

\bibitem[{\textit{{Ingersoll} et~al.}(2017)}]{Ingersoll2017}
{Ingersoll}, A.~P., et~al., {Implications of the ammonia distribution on
  Jupiter from 1 to 100 bars as measured by the Juno microwave radiometer},
  \textit{Geophys. Res. Lett.}, \textit{44}(15), 7676--7685,
  \doi{10.1002/2017GL074277}, 2017.

\bibitem[{\textit{{Janssen} et~al.}(2017)}]{Janssen2017}
{Janssen}, M.~A., et~al., {MWR: Microwave Radiometer for the Juno Mission to
  Jupiter}, \textit{Space Sci. Rev.}, \textit{213}(1-4), 139--185,
  \doi{10.1007/s11214-017-0349-5}, 2017.

\bibitem[{\textit{{Kaspi}}(2013)}]{Kaspi2013a}
{Kaspi}, Y., Inferring the depth of the zonal jets on {J}upiter and {S}aturn
  from odd gravity harmonics, \textit{Geophys. Res. Lett.}, \textit{40},
  676--680, \doi{10.1029/2009GL041385}, 2013.

\bibitem[{\textit{{Kaspi} et~al.}(2009)\textit{{Kaspi}, {Flierl}, and
  {Showman}}}]{Kaspi2009}
{Kaspi}, Y., G.~R. {Flierl}, and A.~P. {Showman}, {The deep wind structure of
  the giant planets: Results from an anelastic general circulation model},
  \textit{Icarus}, \textit{202}(2), 525--542,
  \doi{10.1016/j.icarus.2009.03.026}, 2009.

\bibitem[{\textit{{Kaspi} et~al.}(2010)\textit{{Kaspi}, {Hubbard}, {Showman},
  and {Flierl}}}]{Kaspi2010a}
{Kaspi}, Y., W.~B. {Hubbard}, A.~P. {Showman}, and G.~R. {Flierl},
  Gravitational signature of {J}upiter's internal dynamics, \textit{Geophys.
  Res. Lett.}, \textit{37}, L01,204, \doi{10.1029/2009GL041385}, 2010.

\bibitem[{\textit{{Kaspi} et~al.}(2020)\textit{{Kaspi}, {Galanti}, {Showman},
  {Stevenson}, {Guillot}, {Iess}, and {Bolton}}}]{Kaspi2020}
{Kaspi}, Y., E.~{Galanti}, A.~P. {Showman}, D.~J. {Stevenson}, T.~{Guillot},
  L.~{Iess}, and S.~J. {Bolton}, {Comparison of the Deep Atmospheric Dynamics
  of Jupiter and Saturn in Light of the Juno and Cassini Gravity Measurements},
  \textit{Space Sci. Rev.}, \textit{216}(5), 84,
  \doi{10.1007/s11214-020-00705-7}, 2020.

\bibitem[{\textit{{Kaspi} et~al.}(2018)}]{Kaspi2018}
{Kaspi}, Y., et~al., Jupiter's atmospheric jet streams extend thousands of
  kilometres deep, \textit{Nature}, \textit{555}, 223--226,
  \doi{10.1038/nature25793}, 2018.

\bibitem[{\textit{{Kong} et~al.}(2018)\textit{{Kong}, {Zhang}, {Schubert}, and
  {Anderson}}}]{Kong2018}
{Kong}, D., K.~{Zhang}, G.~{Schubert}, and J.~D. {Anderson}, Origin of
  {J}upiter's cloud-level zonal winds remains a puzzle even after {J}uno,
  \textit{Proc. Natl. Acad. Sci. U.S.A.}, \textit{115}(34), 8499--8504,
  \doi{10.1073/pnas.1805927115}, 2018.

\bibitem[{\textit{Li et~al.}(2017)}]{Li2017}
Li, C., et~al., The distribution of ammonia on {J}upiter from a preliminary
  inversion of {J}uno microwave radiometer data, \textit{Geophys. Res. Lett.},
  \textit{44}(11), 5317--5325, \doi{10.1002/2017GL073159}, 2017.

\bibitem[{\textit{{Li} et~al.}(2020)}]{Li2020}
{Li}, C., et~al., {The water abundance in Jupiter's equatorial zone},
  \textit{Nature Astronomy}, \textit{4}, 609--616,
  \doi{10.1038/s41550-020-1009-3}, 2020.

\bibitem[{\textit{{Li} et~al.}(2006)\textit{{Li}, {Ingersoll}, {Vasavada},
  {Simon-Miller}, {Del Genio}, {Ewald}, {Porco}, and {West}}}]{Li2006a}
{Li}, L., A.~P. {Ingersoll}, A.~R. {Vasavada}, A.~A. {Simon-Miller}, A.~D. {Del
  Genio}, S.~P. {Ewald}, C.~C. {Porco}, and R.~A. {West}, {Vertical wind shear
  on Jupiter from Cassini images}, \textit{J. Geophys. Res. (Planets)},
  \textit{111}(E4), E04004, \doi{10.1029/2005JE002556}, 2006.

\bibitem[{\textit{{Liu} et~al.}(2008)\textit{{Liu}, {Goldreich}, and
  {Stevenson}}}]{Liu2008}
{Liu}, J., P.~M. {Goldreich}, and D.~J. {Stevenson}, Constraints on deep-seated
  zonal winds inside {J}upiter and {S}aturn, \textit{Icarus}, \textit{196},
  653--664, \doi{10.1016/j.icarus.2007.11.036}, 2008.

\bibitem[{\textit{{Moore} et~al.}(2019)\textit{{Moore}, {Cao}, {Bloxham},
  {Stevenson}, {Connerney}, and {Bolton}}}]{Moore2019}
{Moore}, K.~M., H.~{Cao}, J.~{Bloxham}, D.~J. {Stevenson}, J.~E.~P.
  {Connerney}, and S.~J. {Bolton}, {Time variation of Jupiter's internal
  magnetic field consistent with zonal wind advection}, \textit{Nature
  Astronomy}, \textit{3}, 730--735, \doi{10.1038/s41550-019-0772-5}, 2019.

\bibitem[{\textit{{Orton} et~al.}(1998)}]{Orton1998}
{Orton}, G.~S., et~al., Characteristics of the {G}alileo probe entry site from
  earth-based remote sensing observations, \textit{J. Geophys. Res.},
  \textit{103}, 22,791--22,814, \doi{10.1029/98JE02380}, 1998.

\bibitem[{\textit{Tollefson et~al.}(2017)}]{Tollefson2017}
Tollefson, J., et~al., Changes in {J}upiter's zonal wind profile preceding and
  during the {J}uno mission, \textit{Icarus}, \textit{296}, 163--178,
  \doi{10.1016/j.icarus.2017.06.007}, 2017.

\bibitem[{\textit{{Wicht} et~al.}(2019)\textit{{Wicht}, {Gastine}, {Duarte},
  and {Dietrich}}}]{Wicht2019a}
{Wicht}, J., T.~{Gastine}, L.~D.~V. {Duarte}, and W.~{Dietrich}, {Dynamo action
  of the zonal winds in {J}upiter}, \textit{Astron. and Astrophys.},
  \textit{629}, A125, \doi{10.1051/0004-6361/201935682}, 2019.



\end{thebibliography}
\end{document}